\documentclass[%
 reprint,
 amsmath,amssymb,
 aps,
 prl,
]{revtex4-2}

\usepackage{graphicx}
\usepackage{dcolumn}
\usepackage{bm}
\usepackage{float}
\usepackage{mathtools} 
\usepackage{physics}
\usepackage{subfiles}
\usepackage{appendix} 

\def\probrigidity{P}
\def\probsinglenoderigid{P_N}
\def\probrods{p_{r}}
\def\probcables{p_{c}}
\def\probstruts{p_{s}}
\def\nonlinearratio{f_c}
\def\probNL{p_t}

\def\extension{\textbf{e}}

\def\elementmatrix{\textbf{K}}

\begin{document}

\preprint{APS/123-QED}

\title{Rigidity of generic random tensegrity structures}

\author{Vishal Sudhakar} 
\affiliation{School of Physics, Georgia Institute of Technology, Atlanta, GA 30332}
\author{James P. McInerney} 
\affiliation{Department of Physics, University of Michigan, Ann Arbor, MI 48109}
\author{D. Zeb Rocklin}
\affiliation{School of Physics, Georgia Institute of Technology, Atlanta, GA 30332}
\author{William Stephenson}
\affiliation{Department of Physics, University of Michigan, Ann Arbor, MI 48109}

\date{\today}

\begin{abstract}
Many mechanical structures, both engineered and biological, combine heavy rigid elements such as bones and beams with lightweight flexible ones such as cables and membranes. 
These are referred to as tensegrities, reflecting that cables can only support extensile tension.
We model such systems via simulations of depleted triangular lattices in which we minimize the energies of tensegrities subject to strained boundary conditions.
When there are equal numbers of cables and struts (which support only compressive tension), a cable and a strut together each contribute as much toward rigidity as a rod, with the two contributions being equal in the case of shear strain.
Due to the highly nonaffine deformations at the rigidity transitions, the contribution of a cable (strut) can be significant even under global compression (dilation) despite a cable's inability to resist local compression.
Further, we find that when neighboring elements tend to point away from one another, as is common in real systems, cables interact significantly more strongly with other cables than do cables with struts in supporting stress.
These phenomena shed new light on a variety of realistic, disordered systems at the threshold of mechanical stability.
\end{abstract}

\maketitle
Many structures observed around the world consist of two categories of elements: linear and nonlinear. 
These systems belong to the class of mechanical structures known as tensegrity structures~\cite{roth1981tensegrity}. 
Tensegrity structures, characterized by a balance of tensile and compressive forces within a network of rods, cables, and struts, have remarkable mechanical properties and potential applications in various fields ranging from architecture to biology~\cite{fu2005structural, zhang2015constructing, carpentieri2017dynamics, bauer2021tensegrity}. 
Tensegrity elements comprise one type of linear element (Hookean springs or rods) and two types of nonlinear elements (cables and struts). 
In a purely linearly constrained system, the emergence of bulk rigidity is predicted by Calladine's extension of Maxwell's count~\cite{calladine1978buckminster}. 
The rigidity percolation transition determines a system's ability to support external loads, which has previously been studied in a generic network under central force using a combinatorial algorithm~\cite{zhang2015rigidity}. 
In one such system, the topology resembles a triangular lattice with locally distorted bond lengths and bond angles, also known as a depleted triangular lattice~\cite{jacobs1995generic}. 
These systems are composed only of rigid rods, which can neither be stretched nor compressed without an energy cost, posing a linear problem.
When tensegrity elements, such as cables or struts, are introduced into the system, the problem naturally becomes nonlinear. 
Cables cannot support compression and struts cannot support extension. 
These actions of the tensegrity elements do not require energy, which forms one-way constraints. 
The nonlinear and geometric nature of the problem makes analytical computations challenging~\cite{connelly1992stability}. 
In biological cells, nonlinear tensegrity structures manifest as microtubules and actin microfilaments. 
The configurations of these structures within the cytoskeleton of cells influence the cell shape and mechanical stability. 
Furthermore, they play a crucial role in driving cellular mechanotransduction, the conversion of physical forces into biological signals~\cite{ingber2014tensegrity}. Previously, Stephenson et al.~\cite{stephenson2023rigidity} illustrated that when nonlinear elements are treated as half the linear elements, the rigidity transition occurs at the Maxwell point in a generic square lattice system.
While disordered systems~\cite{feng2016nonlinear, zaiser2023disordered} and tensegrity structures~\cite{snelson1996, skelton2009tensegrity, rimoli2017mechanical} have been individually explored, their concurrent study poses a nontrivial analysis.
In this paper, we demonstrate that in a depleted triangular lattice, under symmetric conditions, nonlinear elements do count as half of the linear elements towards rigidity. 
We further find that nonaffine deformations enable singular types of nonlinear elements (ie, only cables or only struts), along with linear elements, to support stress even under unfavorable metric strains. 
Moreover, under neutral shear, singular types of nonlinear elements present in the system contribute more towards the system's rigidity than a mixture of both types.
\begin{figure*}
  \includegraphics[width=\textwidth]{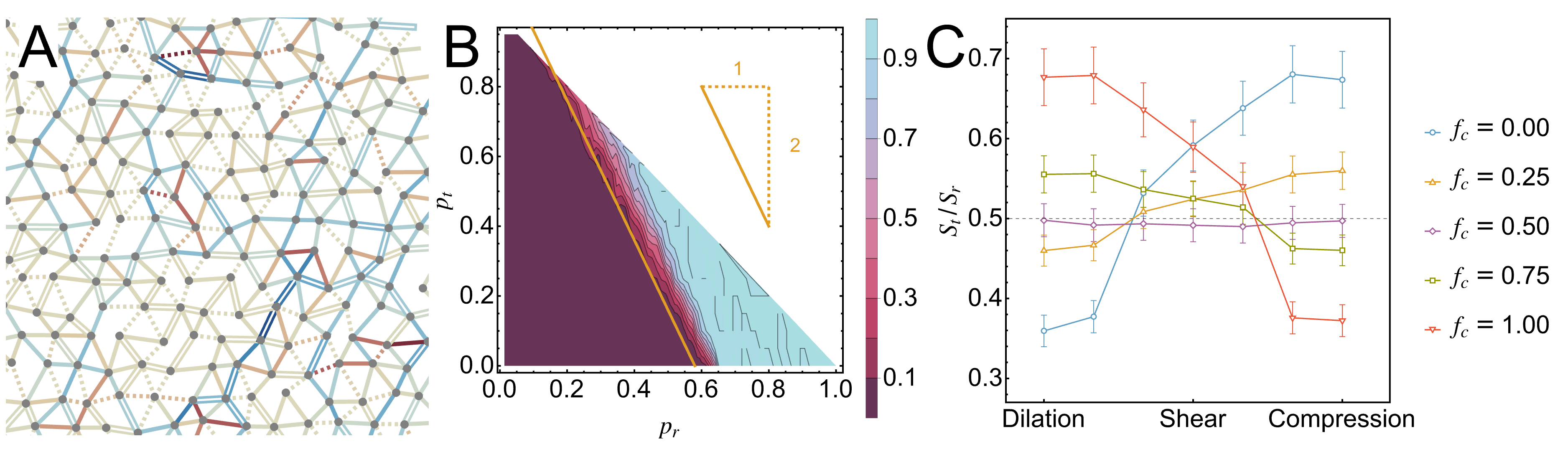}
  \caption{(A) A force-balanced configuration of the depleted triangular lattice with rods (solid lines), cables (dashed lines), and struts (hollow lines) subject to periodic boundary conditions. The colors from red to green to blue depict positive to zero to negative tensions in the elements. 
  (B) The probability of rigidity as a function of the probability of rods, $\probrods$, and the probability of nonlinear elements, $\probNL$, with the 50-50 cable-to-strut fraction ($\nonlinearratio = 0.5$) under a shear strain applied to the system. The contour linear aligns with the theory (orange line), indicating that nonlinear elements support rigidity half as much as linear elements.
  (C) The efficacy of non-linear elements over the efficacy of linear elements for various values of $\nonlinearratio$ and different types of strains. When an equal fraction of cables and struts is present ($\nonlinearratio = 0.5$), nonlinear elements contribute as half of linear elements towards rigidity ($S_t / S_r = 0.5$). Under compression (dilation), cables (struts) non-intuitively contribute about $37.2 \% \pm 1.3 \%$ ($35.9 \% \pm 1.5 \%$) as rods towards the rigidity of the system. We also find that on average nonlinear elements of one type (only cables or only struts), along with linear elements, can withstand all neutral shear types of strains better than an equal mixture of cables and struts.} 
  \label{fig:figure_one}
\end{figure*}

We consider periodic, depleted triangular lattices of mechanical elements such as that shown in Fig.~\ref{fig:figure_one}A, in which individual nodes are randomly displaced to prevent straight one-dimensional lines of force-bearing elements.
An $ n \times n$ system has $3 n^2$ sites where elements can be placed between neighboring nodes. 
At each, a harmonic, linear spring is present with probability $\probrods$, while a \emph{tensegrity} element is present with probability $\probNL$. 
Of the latter, a fraction $\nonlinearratio$ are cables, which generate tension only under extensions, while the rest are struts which generate tension only under compression~\cite{connelly1992stability}. 
Consequently, the energy of the system is
\begin{align}\label{energy_matrix_form}
    E = \frac{1}{2}\extension^{T}\cdot\elementmatrix(\extension)\cdot \extension
\end{align}

\noindent where $\extension$ is the vector of bond extensions. 
$\elementmatrix$ is a diagonal matrix with components

\begin{align}
    K_{ii}(e_i) = \begin{cases}
        0             & \text{: no element}\\
        1             & \text{: rods}\\
        \Theta(e_i)   & \text{: cables}\\
        \Theta(-e_i)  & \text{: struts}
    \end{cases} 
\end{align}

\noindent where $\Theta(\cdot)$ is the Heaviside step function and for simplicity we assume each element has the same spring constant, which we set to unity.
We subject the system to metric strains (dilation, shear, compression) parameterized by $\phi$, with $\phi = 0$ for pure dilation, $\phi = \pi/2$ for neutral shear, and $\phi = \pi$ for pure compression. 
The system undergoes additional relaxations to lower its energy and balance forces, as detailed in the SI. Energies are minimized numerically in \emph{Mathematica}.

We obtain the probability that the system rigidly resists strain $\phi$, $P_\phi(\probrods,\probNL,\nonlinearratio)$, from the fraction of randomly generated systems that have positive energy, as shown in Fig.~\ref{fig:figure_one}B for the case of shear strain and equal number of cables and struts $\nonlinearratio = 0.5$. 
Because a rod is equivalent to the superposition of a cable and a strut connecting the same two nodes, a mean field picture in which the positions and orientations of elements are ignored implies that a rod must contribute towards rigidity as much as a cable plus a strut. 
Indeed, as can be seen in that figure, the probability is a function of $\probrods + \frac{1}{2}\probNL$ only, meaning that each cable or strut contributes half as much towards rigidity as does a rod, as previously found in a more restricted set of random tensegrities~\cite{stephenson2023rigidity}. 
More generally, we conjecture that for a given form of strain and a given $\nonlinearratio$, a nonlinear element might contribute more or less to rigidity than in the mean field picture, such that the probability function takes the form

\begin{align}
    P_\phi(\probrods,\probNL,\nonlinearratio) = P_\phi(\probrods + \frac{S_t}{S_r} \probNL, \nonlinearratio),
\end{align}

\noindent where $S_t / S_r$, a function of both strain and $\nonlinearratio$, describes the contribution of the nonlinear element towards rigidity, relative to a rod. 
We refer to the quantities $S_t \equiv \partial P_\phi/\partial \probNL$ and $S_r \equiv \partial P_\phi/\partial \probrods$ as the \emph{efficacies} of nonlinear elements and linear elements, respectively. 
Even without knowing the particular form of the probability of rigidity, the efficacy of tensegrity elements compared to rods can be measured from simulation data by calculating $(\partial P_\phi/\partial \probNL)/(\partial P_\phi/\partial \probrods)$. 

We find that when there is an equal amount of cables and struts in the system ($\nonlinearratio = 0.5$), as shown in Fig.~\ref{fig:figure_one}C, the nonlinear elements always count as half the linear elements in all types of metric strain, as Stephenson et al.~\cite{stephenson2023rigidity} found for ideal square lattices. 
Interestingly, when we deviate away from this symmetric case, varying $\nonlinearratio$, we see that nonlinear elements can withstand external forces more than one might expect. 
For example, in the case of only cables and rods present in the system subject to compression, we would naively expect that cables will not contribute anything to the system's rigidity because they are unable to resist compression.
However, we find that in such system configurations, cables are $37\%$ as effective as rods. 
Moreover, cables under dilation, where they are naturally effective, count as much as $67\%$ as much as rods, far more than the expected $50\%$.
This implies that in disordered tensegrity structures, using a single type of nonlinear element, such as cytoskeletons in cells~\cite{ingber2008tensegrity3}, can contribute to rigidity, though its effectiveness varies depending on the types of strains.
\begin{figure}
    \centering
    \includegraphics[width=\columnwidth]{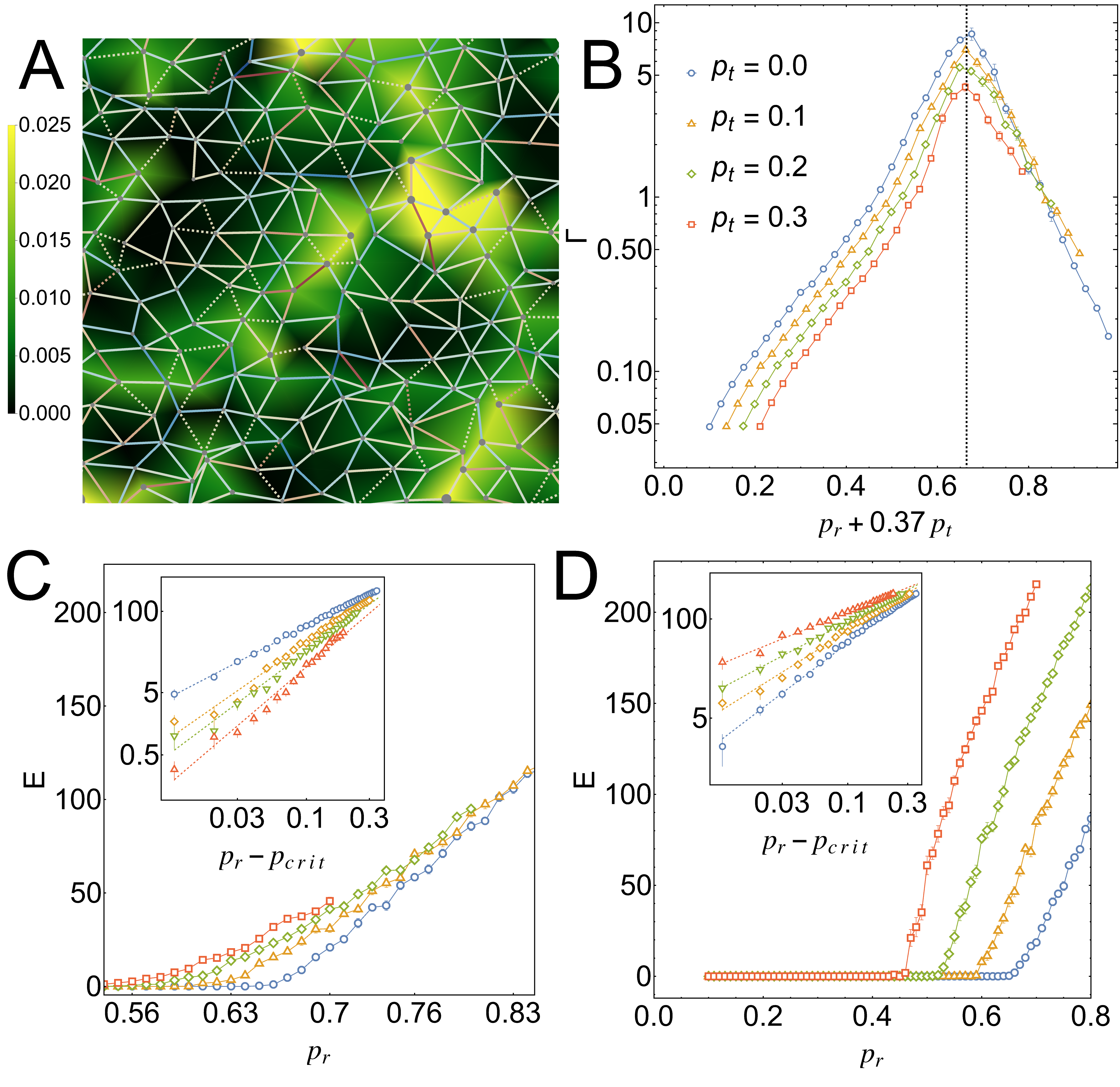}
    \caption{(A) A system of rods and cables (without struts) undergoing compression, with the amount of nonaffine displacement depicted with the intensity of background color. (B) Deformations can be highly nonaffine, as denoted by $\Gamma$, reaching peaks at the rigidity transition (dashed line) as the densities of rods ($\probrods$) and cables ($\probNL$) increase.
    (C) The energy of the system under compression begins to increase at the rigidity transition, which occurs with fewer rods when cables are present. The inset shows the scaling near the transition.
    (D) The energy of the same system under dilation rather than compression. Here, the presence of cables leads to an earlier transition, revealing a finite regime in which systems resist dilation but not compression.}
    \label{fig:figure_two}
\end{figure}

Unfavorable strains are deformations where the nonlinear elements are usually ineffective, e.g. cables (struts) under compression (dilation).
The unique behavior of the nonlinear elements under unfavorable strains is a consequence of the nonaffine displacements of the nodes in the system.
We characterize the nonaffinity by measuring the nonaffinity parameter~\cite{broedersz2011criticality} defined as
\begin{align}
    \Gamma = \frac{1}{l^2\gamma^2n^2}\sum_{i}^{n^2} (u^{i}_{\textrm{na}})^2,
\end{align}
where $l$ is the average bond length, $\gamma$ is the magnitude of the metric strain, and $u_{\textrm{na}}^i$ is the nonaffine displacement of node $i$.
In Fig.~\ref{fig:figure_two}B, we plot the nonaffinity parameter for a system with only cables and rods under compression. 
As the system reaches the transition point, the nonaffinity is at its maximum and about $37\%$ of the cables are under positive tension while, globally, the system is being compressed. 
This allows the cables in the system to withstand external forces, hence contributing partially towards rigidity. 
This characteristic of the system can also be observed by looking at the energy of the system near the transition point (Fig.~\ref{fig:figure_two}C) where the transition to rigidity occurs earlier when more cables are present in the system. 
After transition, as the nonaffinity decreases, cables disengage and the energies of the systems become independent of the number of cables.
Fig.~\ref{fig:figure_two}C's inset is a log-log plot of the same energies showing approximate power-law behavior (dashed lines).
We can observe that the slopes for the curves with cables in the system ($\probNL > 0$) are greater than those with only rods ($\probNL = 0$), 
indicating that the transition occurs more sharply under compression when cables rather than only rods are present.
In Fig.~\ref{fig:figure_two}D, we have the same systems of cables and rods under dilation rather than compression.
Again, the presence of cables leads to an earlier transition point, as even under locally nonaffine deformation the cables are sometimes engaged by the global dilation. The cables smooth the transition under dilation, in contrast with the sharpening observed under compression.
However, as rods are added, the system's deformation becomes increasingly affine, so that cables continue to contribute to the energy far above the transition point, in contrast with the system under compression.

\begin{figure}
  \includegraphics[width=\columnwidth]{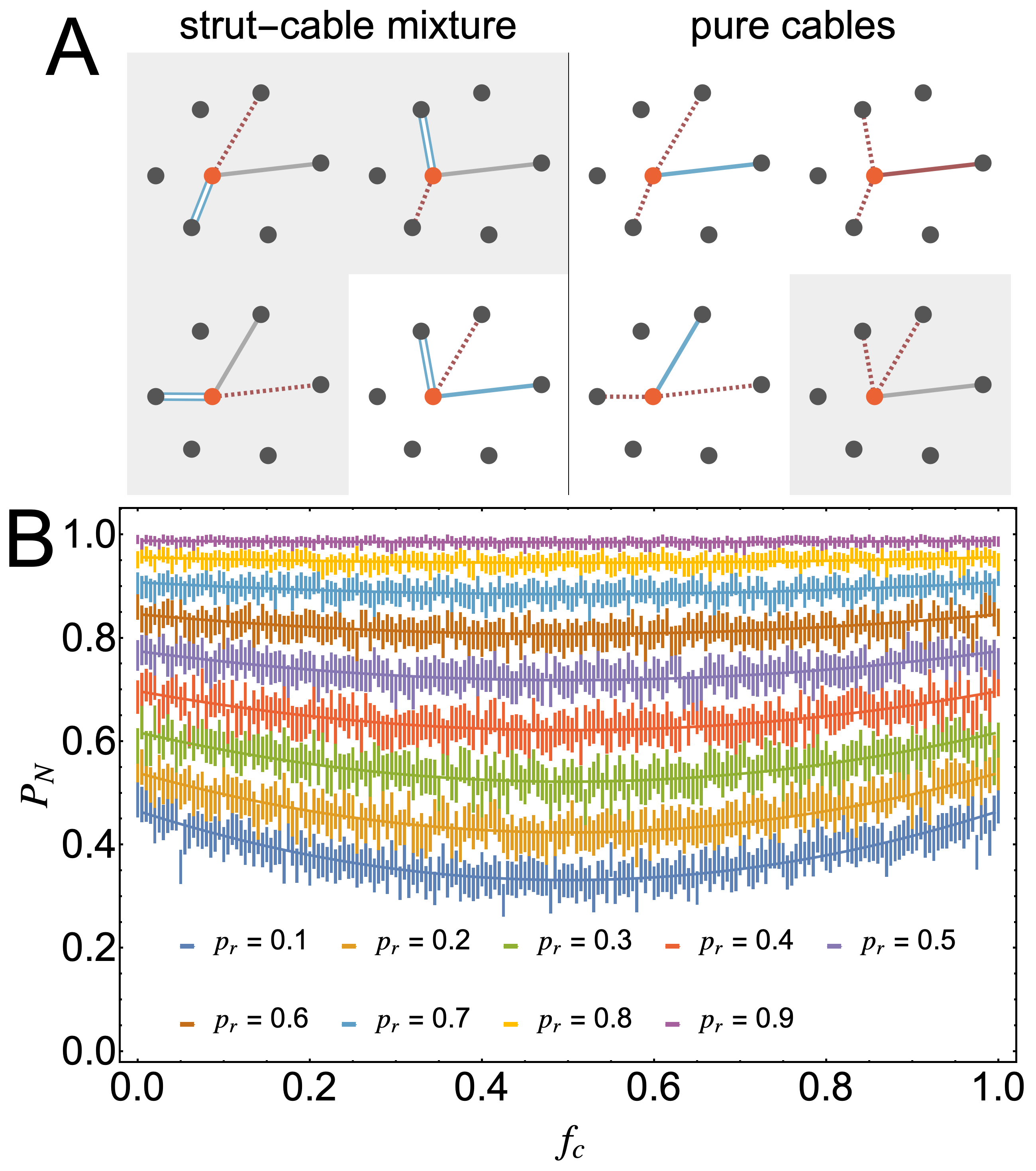}
  \caption{(A) An illustration of single three coordinated configurations having strictly cables (on the right) results in more force-balanced configurations, rather than having a mixture of cables and struts (on the left). The non-force-balanced configurations are depicted with gray backgrounds, indicating that it is impossible to assign a force to the third element (in gray) in a way that makes the configuration force-balanced. The opposite case may be true, where a mixture of cables and struts can form a force-balanced configuration while strictly cables or struts cannot (bottom right squares). However, when considering all possible three-element configurations, there are more force-balanced configurations with only cables or struts than with a mixture of them.
  (B) The probability, $\probsinglenoderigid$, that a single three-coordinated node is rigid as a function of $\nonlinearratio$ for different values of $\probrods$. The analytical solutions are shown as solid lines with error bars generated from simulation data. The probability of a single node being rigid is greatest when the system has cables (struts) present exclusively, indicating its enhanced stress resistance compared to a mixture of cables and struts present in the system. Thus, tensegrity systems with nonlinear elements of one type can withstand external forces better.} 
  \label{fig:figure_three}
\end{figure}

We now consider the difference between cable-cable and cable-strut interactions in a system  undergoing compression along one axis and dilation about the other, a pure shear strain which neither dilates nor compresses the system.
We find that systems with only one type of tensegrity element are more likely to support stress than those with a mixture of cables and struts.
This is because, as shown in Fig.~\ref{fig:figure_three}A, elements emanating from a single vertex tend to point in opposing directions. Two elements with positive tensions (e.g., two cables) are then more likely to balance forces than if one has negative tension (e.g., a cable and a rod).
We illustrate this in the figure with three-coordinated vertices, in which three of the four cases support stress when two cables are present, whereas only one of the cases supports stress when a cable and strut are present. 
The probabilities of such a three-coordinated vertex supporting stress are shown in Fig.~\ref{fig:figure_three}B, with those probabilities being maximal for all-cable and all-strut systems and minimal for an even mixture of cables and struts, with details of the simulation and analytic model shown in the Supplementary Material.
This effect is also evident in Fig.~\ref{fig:figure_one}C, in which (averaged over all types of strain) nonlinear elements contribute more to rigidity under shear strains when $\nonlinearratio > 0.5$ and $\nonlinearratio < 0.5$ than at $\nonlinearratio = 0.5$.
\begin{figure}
  \includegraphics[width=\columnwidth]{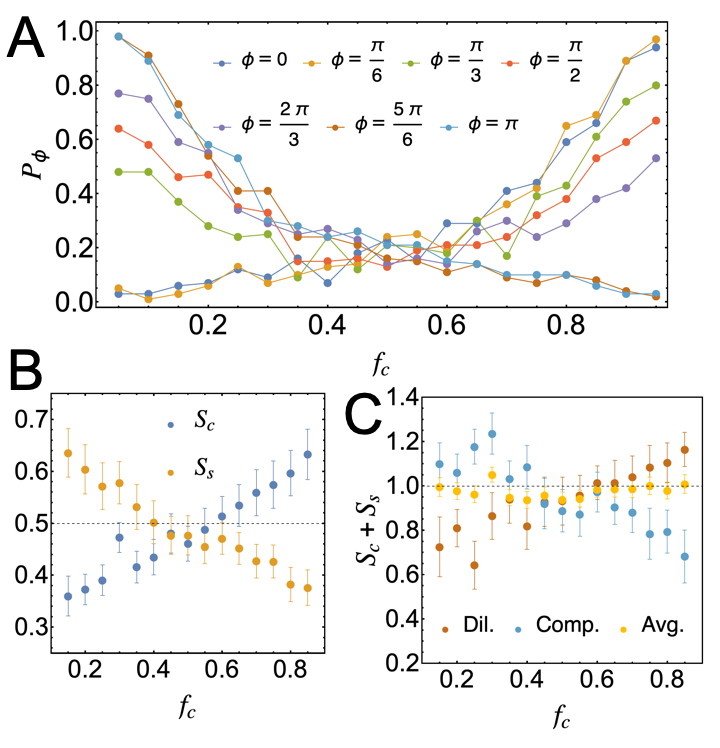}
  \caption{(A) The probability of rigidity as a function of cable fraction, $\nonlinearratio$, with the type of strain parameterized by $\phi$, for which 0 corresponds to dilation, $\pi/2$ to shear and $\pi$ to compression. Here, the fraction of bonds present, $\probrods$, is .3 and the combined fraction of cables and struts, $\probNL$ is .6, and cables (struts) contribute more to rigidity under dilation (compression). (B) The efficacy, or contribution to rigidity relative to that of a rod, of cables (struts), $S_{c(s)}$, averaged over all metric strains, in which the presence of other cables (struts) increases the efficacy of an additional cable (strut). (C) The combined efficacy of a cable and a strut is independent of $\nonlinearratio$ for averaged strains, but increases (decreases) with $\nonlinearratio$ for dilation (compression).} 
  \label{fig:figure_four}
\end{figure}

Finally, we examine the complex interplay between the types of nonlinear elements present and the types of strain applied, while fixing parameters $\probrods = 0.3$ and $\probNL = 0.6$.
In Fig.~\ref{fig:figure_four}A, we observe the expected behavior that the probability of rigidity increases (decreases) as we increase $\nonlinearratio$ for more dilational (compressional) strains. 
Consequently, in Fig.~\ref{fig:figure_four}B the efficacies of cables (struts) averaged over all strains $S_{c(s)}$ increase (decrease) with cable fraction.
As shown in Fig. \ref{fig:figure_four}C, the combined efficacies of a cable and a strut, $S_{c} + S_{s}$, add up to that of a rod when averaged over all strain, irrespective of $\nonlinearratio$ indicating that cables and struts each contribute half as much as rods. 
However, for strictly dilational (compressive) strains, we observe that this combined efficacy is greater than that of a rod for $\nonlinearratio > 0.5$ ($\nonlinearratio < 0.5$). 
This final result confirms the mean-field result of Fig.~\ref{fig:figure_three}, that the microstructure favors the efficacy of strut-strut or cable-cable interactions over strut-cable interactions in resisting particular strains.

In many realistic disordered tensegrity systems, nonlinear elements play an important role in the mechanical stability of the system and other mechanical functions~\cite{ingber2003tensegrity1, ingber2003tensegrity2}. 
In addition to benefits like reduced mass, tensegrity structures enable mechanism design using inequality constraints, yielding a two-dimensional, disk-like configuration space rather than a one-dimensional, circle-like space characteristic of purely linear elements.
We show that in systems with even mixtures of cables and struts, along with rods subject to all types of strain (dilation to shear to compression), each nonlinear element counts toward rigidity with half the efficacy of a rod, which can itself be thought of as the superposition of a cable and a strut.
However, when a system of rods and cables is subject to compression (or a system of rods and struts to dilation) the nonlinear elements are, on average, $37\%$ as effective as rods, due to nonaffinity resulting in patches of local dilation even with system under global compression (and vice versa).
Even beyond this, the orientational correlations induced by the microstructure, allow pairs of cables or pairs of struts to resist certain types of strain more effectively than a cable-strut pair.
These results incorporating complex, disordered structures and nonlinear tensegrity elements, shed new light on how biological systems across length scales, from cells to musculoskeletal systems, achieve flexibility and the ability to support external loads~\cite{chen1999tensegrity}. 
Finally, these same principles can be used to design new flexible mechanical structures, for which complex structures, nonlinearity and nonaffinity play important roles~\cite{bertoldi2017flexible}.
\paragraph*{Acknowledgments}
We gratefully acknowledge 
financial support from the National Science Foundation CAREER program 
(\#2338492) and from the Army Research Office through the MURI program 
(\#W911NF2210219).
\nocite{*}
\bibliography{citations}

\begin{thebibliography}{24}%
\makeatletter
\providecommand \@ifxundefined [1]{%
 \@ifx{#1\undefined}
}%
\providecommand \@ifnum [1]{%
 \ifnum #1\expandafter \@firstoftwo
 \else \expandafter \@secondoftwo
 \fi
}%
\providecommand \@ifx [1]{%
 \ifx #1\expandafter \@firstoftwo
 \else \expandafter \@secondoftwo
 \fi
}%
\providecommand \natexlab [1]{#1}%
\providecommand \enquote  [1]{``#1''}%
\providecommand \bibnamefont  [1]{#1}%
\providecommand \bibfnamefont [1]{#1}%
\providecommand \citenamefont [1]{#1}%
\providecommand \href@noop [0]{\@secondoftwo}%
\providecommand \href [0]{\begingroup \@sanitize@url \@href}%
\providecommand \@href[1]{\@@startlink{#1}\@@href}%
\providecommand \@@href[1]{\endgroup#1\@@endlink}%
\providecommand \@sanitize@url [0]{\catcode `\\12\catcode `\$12\catcode `\&12\catcode `\#12\catcode `\^12\catcode `\_12\catcode `\%12\relax}%
\providecommand \@@startlink[1]{}%
\providecommand \@@endlink[0]{}%
\providecommand \url  [0]{\begingroup\@sanitize@url \@url }%
\providecommand \@url [1]{\endgroup\@href {#1}{\urlprefix }}%
\providecommand \urlprefix  [0]{URL }%
\providecommand \Eprint [0]{\href }%
\providecommand \doibase [0]{https://doi.org/}%
\providecommand \selectlanguage [0]{\@gobble}%
\providecommand \bibinfo  [0]{\@secondoftwo}%
\providecommand \bibfield  [0]{\@secondoftwo}%
\providecommand \translation [1]{[#1]}%
\providecommand \BibitemOpen [0]{}%
\providecommand \bibitemStop [0]{}%
\providecommand \bibitemNoStop [0]{.\EOS\space}%
\providecommand \EOS [0]{\spacefactor3000\relax}%
\providecommand \BibitemShut  [1]{\csname bibitem#1\endcsname}%
\let\auto@bib@innerbib\@empty
\bibitem [{\citenamefont {Roth}\ and\ \citenamefont {Whiteley}(1981)}]{roth1981tensegrity}%
  \BibitemOpen
  \bibfield  {author} {\bibinfo {author} {\bibfnamefont {B.}~\bibnamefont {Roth}}\ and\ \bibinfo {author} {\bibfnamefont {W.}~\bibnamefont {Whiteley}},\ }\bibfield  {title} {\bibinfo {title} {Tensegrity frameworks},\ }\href@noop {} {\bibfield  {journal} {\bibinfo  {journal} {Transactions of the American Mathematical Society}\ }\textbf {\bibinfo {volume} {265}},\ \bibinfo {pages} {419} (\bibinfo {year} {1981})}\BibitemShut {NoStop}%
\bibitem [{\citenamefont {Fu}(2005)}]{fu2005structural}%
  \BibitemOpen
  \bibfield  {author} {\bibinfo {author} {\bibfnamefont {F.}~\bibnamefont {Fu}},\ }\bibfield  {title} {\bibinfo {title} {Structural behavior and design methods of tensegrity domes},\ }\href@noop {} {\bibfield  {journal} {\bibinfo  {journal} {Journal of Constructional Steel Research}\ }\textbf {\bibinfo {volume} {61}},\ \bibinfo {pages} {23} (\bibinfo {year} {2005})}\BibitemShut {NoStop}%
\bibitem [{\citenamefont {Zhang}\ \emph {et~al.}(2015{\natexlab{a}})\citenamefont {Zhang}, \citenamefont {Zhao},\ and\ \citenamefont {Feng}}]{zhang2015constructing}%
  \BibitemOpen
  \bibfield  {author} {\bibinfo {author} {\bibfnamefont {L.-Y.}\ \bibnamefont {Zhang}}, \bibinfo {author} {\bibfnamefont {H.-P.}\ \bibnamefont {Zhao}},\ and\ \bibinfo {author} {\bibfnamefont {X.-Q.}\ \bibnamefont {Feng}},\ }\bibfield  {title} {\bibinfo {title} {Constructing large-scale tensegrity structures with bar--bar connection using prismatic elementary cells},\ }\href@noop {} {\bibfield  {journal} {\bibinfo  {journal} {Archive of Applied Mechanics}\ }\textbf {\bibinfo {volume} {85}},\ \bibinfo {pages} {383} (\bibinfo {year} {2015}{\natexlab{a}})}\BibitemShut {NoStop}%
\bibitem [{\citenamefont {Carpentieri}\ and\ \citenamefont {Skelton}(2017)}]{carpentieri2017dynamics}%
  \BibitemOpen
  \bibfield  {author} {\bibinfo {author} {\bibfnamefont {G.}~\bibnamefont {Carpentieri}}\ and\ \bibinfo {author} {\bibfnamefont {R.}~\bibnamefont {Skelton}},\ }\bibfield  {title} {\bibinfo {title} {On the dynamics of tensegrity bridges},\ }\href@noop {} {\bibfield  {journal} {\bibinfo  {journal} {J. Aerosp. Eng. Mech}\ }\textbf {\bibinfo {volume} {1}},\ \bibinfo {pages} {48} (\bibinfo {year} {2017})}\BibitemShut {NoStop}%
\bibitem [{\citenamefont {Bauer}\ \emph {et~al.}(2021)\citenamefont {Bauer}, \citenamefont {Kraus}, \citenamefont {Crook}, \citenamefont {Rimoli},\ and\ \citenamefont {Valdevit}}]{bauer2021tensegrity}%
  \BibitemOpen
  \bibfield  {author} {\bibinfo {author} {\bibfnamefont {J.}~\bibnamefont {Bauer}}, \bibinfo {author} {\bibfnamefont {J.~A.}\ \bibnamefont {Kraus}}, \bibinfo {author} {\bibfnamefont {C.}~\bibnamefont {Crook}}, \bibinfo {author} {\bibfnamefont {J.~J.}\ \bibnamefont {Rimoli}},\ and\ \bibinfo {author} {\bibfnamefont {L.}~\bibnamefont {Valdevit}},\ }\bibfield  {title} {\bibinfo {title} {Tensegrity metamaterials: toward failure-resistant engineering systems through delocalized deformation},\ }\href@noop {} {\bibfield  {journal} {\bibinfo  {journal} {Advanced Materials}\ }\textbf {\bibinfo {volume} {33}},\ \bibinfo {pages} {2005647} (\bibinfo {year} {2021})}\BibitemShut {NoStop}%
\bibitem [{\citenamefont {Calladine}(1978)}]{calladine1978buckminster}%
  \BibitemOpen
  \bibfield  {author} {\bibinfo {author} {\bibfnamefont {C.~R.}\ \bibnamefont {Calladine}},\ }\bibfield  {title} {\bibinfo {title} {Buckminster fuller's “tensegrity” structures and clerk maxwell's rules for the construction of stiff frames},\ }\href@noop {} {\bibfield  {journal} {\bibinfo  {journal} {International journal of solids and structures}\ }\textbf {\bibinfo {volume} {14}},\ \bibinfo {pages} {161} (\bibinfo {year} {1978})}\BibitemShut {NoStop}%
\bibitem [{\citenamefont {Zhang}\ \emph {et~al.}(2015{\natexlab{b}})\citenamefont {Zhang}, \citenamefont {Rocklin}, \citenamefont {Chen},\ and\ \citenamefont {Mao}}]{zhang2015rigidity}%
  \BibitemOpen
  \bibfield  {author} {\bibinfo {author} {\bibfnamefont {L.}~\bibnamefont {Zhang}}, \bibinfo {author} {\bibfnamefont {D.~Z.}\ \bibnamefont {Rocklin}}, \bibinfo {author} {\bibfnamefont {B.~G.-g.}\ \bibnamefont {Chen}},\ and\ \bibinfo {author} {\bibfnamefont {X.}~\bibnamefont {Mao}},\ }\bibfield  {title} {\bibinfo {title} {Rigidity percolation by next-nearest-neighbor bonds on generic and regular isostatic lattices},\ }\href@noop {} {\bibfield  {journal} {\bibinfo  {journal} {Physical Review E}\ }\textbf {\bibinfo {volume} {91}},\ \bibinfo {pages} {032124} (\bibinfo {year} {2015}{\natexlab{b}})}\BibitemShut {NoStop}%
\bibitem [{\citenamefont {Jacobs}\ and\ \citenamefont {Thorpe}(1995)}]{jacobs1995generic}%
  \BibitemOpen
  \bibfield  {author} {\bibinfo {author} {\bibfnamefont {D.~J.}\ \bibnamefont {Jacobs}}\ and\ \bibinfo {author} {\bibfnamefont {M.~F.}\ \bibnamefont {Thorpe}},\ }\bibfield  {title} {\bibinfo {title} {Generic rigidity percolation: the pebble game},\ }\href@noop {} {\bibfield  {journal} {\bibinfo  {journal} {Physical review letters}\ }\textbf {\bibinfo {volume} {75}},\ \bibinfo {pages} {4051} (\bibinfo {year} {1995})}\BibitemShut {NoStop}%
\bibitem [{\citenamefont {Connelly}\ and\ \citenamefont {Whiteley}(1992)}]{connelly1992stability}%
  \BibitemOpen
  \bibfield  {author} {\bibinfo {author} {\bibfnamefont {R.}~\bibnamefont {Connelly}}\ and\ \bibinfo {author} {\bibfnamefont {W.}~\bibnamefont {Whiteley}},\ }\bibfield  {title} {\bibinfo {title} {The stability of tensegrity frameworks},\ }\href@noop {} {\bibfield  {journal} {\bibinfo  {journal} {International Journal of Space Structures}\ }\textbf {\bibinfo {volume} {7}},\ \bibinfo {pages} {153} (\bibinfo {year} {1992})}\BibitemShut {NoStop}%
\bibitem [{\citenamefont {Ingber}\ \emph {et~al.}(2014)\citenamefont {Ingber}, \citenamefont {Wang},\ and\ \citenamefont {Stamenovi{\'c}}}]{ingber2014tensegrity}%
  \BibitemOpen
  \bibfield  {author} {\bibinfo {author} {\bibfnamefont {D.~E.}\ \bibnamefont {Ingber}}, \bibinfo {author} {\bibfnamefont {N.}~\bibnamefont {Wang}},\ and\ \bibinfo {author} {\bibfnamefont {D.}~\bibnamefont {Stamenovi{\'c}}},\ }\bibfield  {title} {\bibinfo {title} {Tensegrity, cellular biophysics, and the mechanics of living systems},\ }\href@noop {} {\bibfield  {journal} {\bibinfo  {journal} {Reports on Progress in Physics}\ }\textbf {\bibinfo {volume} {77}},\ \bibinfo {pages} {046603} (\bibinfo {year} {2014})}\BibitemShut {NoStop}%
\bibitem [{\citenamefont {Stephenson}\ \emph {et~al.}(2023)\citenamefont {Stephenson}, \citenamefont {Sudhakar}, \citenamefont {McInerney}, \citenamefont {Czajkowski},\ and\ \citenamefont {Rocklin}}]{stephenson2023rigidity}%
  \BibitemOpen
  \bibfield  {author} {\bibinfo {author} {\bibfnamefont {W.}~\bibnamefont {Stephenson}}, \bibinfo {author} {\bibfnamefont {V.}~\bibnamefont {Sudhakar}}, \bibinfo {author} {\bibfnamefont {J.}~\bibnamefont {McInerney}}, \bibinfo {author} {\bibfnamefont {M.}~\bibnamefont {Czajkowski}},\ and\ \bibinfo {author} {\bibfnamefont {D.~Z.}\ \bibnamefont {Rocklin}},\ }\bibfield  {title} {\bibinfo {title} {Rigidity percolation in a random tensegrity via analytic graph theory},\ }\href@noop {} {\bibfield  {journal} {\bibinfo  {journal} {Proceedings of the National Academy of Sciences}\ }\textbf {\bibinfo {volume} {120}},\ \bibinfo {pages} {e2302536120} (\bibinfo {year} {2023})}\BibitemShut {NoStop}%
\bibitem [{\citenamefont {Feng}\ \emph {et~al.}(2016)\citenamefont {Feng}, \citenamefont {Levine}, \citenamefont {Mao},\ and\ \citenamefont {Sander}}]{feng2016nonlinear}%
  \BibitemOpen
  \bibfield  {author} {\bibinfo {author} {\bibfnamefont {J.}~\bibnamefont {Feng}}, \bibinfo {author} {\bibfnamefont {H.}~\bibnamefont {Levine}}, \bibinfo {author} {\bibfnamefont {X.}~\bibnamefont {Mao}},\ and\ \bibinfo {author} {\bibfnamefont {L.~M.}\ \bibnamefont {Sander}},\ }\bibfield  {title} {\bibinfo {title} {Nonlinear elasticity of disordered fiber networks},\ }\href@noop {} {\bibfield  {journal} {\bibinfo  {journal} {Soft matter}\ }\textbf {\bibinfo {volume} {12}},\ \bibinfo {pages} {1419} (\bibinfo {year} {2016})}\BibitemShut {NoStop}%
\bibitem [{\citenamefont {Zaiser}\ and\ \citenamefont {Zapperi}(2023)}]{zaiser2023disordered}%
  \BibitemOpen
  \bibfield  {author} {\bibinfo {author} {\bibfnamefont {M.}~\bibnamefont {Zaiser}}\ and\ \bibinfo {author} {\bibfnamefont {S.}~\bibnamefont {Zapperi}},\ }\bibfield  {title} {\bibinfo {title} {Disordered mechanical metamaterials},\ }\href@noop {} {\bibfield  {journal} {\bibinfo  {journal} {Nature Reviews Physics}\ }\textbf {\bibinfo {volume} {5}},\ \bibinfo {pages} {679} (\bibinfo {year} {2023})}\BibitemShut {NoStop}%
\bibitem [{\citenamefont {Snelson}(1996)}]{snelson1996}%
  \BibitemOpen
  \bibfield  {author} {\bibinfo {author} {\bibfnamefont {K.}~\bibnamefont {Snelson}},\ }\bibfield  {title} {\bibinfo {title} {Snelson on the tensegrity invention},\ }\href {https://doi.org/10.1177/026635119601-207} {\bibfield  {journal} {\bibinfo  {journal} {International Journal of Space Structures}\ }\textbf {\bibinfo {volume} {11}},\ \bibinfo {pages} {43} (\bibinfo {year} {1996})}\BibitemShut {NoStop}%
\bibitem [{\citenamefont {Skelton}\ and\ \citenamefont {De~Oliveira}(2009)}]{skelton2009tensegrity}%
  \BibitemOpen
  \bibfield  {author} {\bibinfo {author} {\bibfnamefont {R.~E.}\ \bibnamefont {Skelton}}\ and\ \bibinfo {author} {\bibfnamefont {M.~C.}\ \bibnamefont {De~Oliveira}},\ }\href@noop {} {\emph {\bibinfo {title} {Tensegrity systems}}},\ Vol.~\bibinfo {volume} {1}\ (\bibinfo  {publisher} {Springer},\ \bibinfo {year} {2009})\BibitemShut {NoStop}%
\bibitem [{\citenamefont {Rimoli}\ and\ \citenamefont {Pal}(2017)}]{rimoli2017mechanical}%
  \BibitemOpen
  \bibfield  {author} {\bibinfo {author} {\bibfnamefont {J.~J.}\ \bibnamefont {Rimoli}}\ and\ \bibinfo {author} {\bibfnamefont {R.~K.}\ \bibnamefont {Pal}},\ }\bibfield  {title} {\bibinfo {title} {Mechanical response of 3-dimensional tensegrity lattices},\ }\href@noop {} {\bibfield  {journal} {\bibinfo  {journal} {Composites Part B: Engineering}\ }\textbf {\bibinfo {volume} {115}},\ \bibinfo {pages} {30} (\bibinfo {year} {2017})}\BibitemShut {NoStop}%
\bibitem [{\citenamefont {Ingber}(2008)}]{ingber2008tensegrity3}%
  \BibitemOpen
  \bibfield  {author} {\bibinfo {author} {\bibfnamefont {D.~E.}\ \bibnamefont {Ingber}},\ }\bibfield  {title} {\bibinfo {title} {Tensegrity-based mechanosensing from macro to micro},\ }\href {https://doi.org/https://doi.org/10.1016/j.pbiomolbio.2008.02.005} {\bibfield  {journal} {\bibinfo  {journal} {Progress in Biophysics and Molecular Biology}\ }\textbf {\bibinfo {volume} {97}},\ \bibinfo {pages} {163} (\bibinfo {year} {2008})},\ \bibinfo {note} {life and Mechanosensitivity}\BibitemShut {NoStop}%
\bibitem [{\citenamefont {Broedersz}\ \emph {et~al.}(2011)\citenamefont {Broedersz}, \citenamefont {Mao}, \citenamefont {Lubensky},\ and\ \citenamefont {MacKintosh}}]{broedersz2011criticality}%
  \BibitemOpen
  \bibfield  {author} {\bibinfo {author} {\bibfnamefont {C.~P.}\ \bibnamefont {Broedersz}}, \bibinfo {author} {\bibfnamefont {X.}~\bibnamefont {Mao}}, \bibinfo {author} {\bibfnamefont {T.~C.}\ \bibnamefont {Lubensky}},\ and\ \bibinfo {author} {\bibfnamefont {F.~C.}\ \bibnamefont {MacKintosh}},\ }\bibfield  {title} {\bibinfo {title} {Criticality and isostaticity in fibre networks},\ }\href@noop {} {\bibfield  {journal} {\bibinfo  {journal} {Nature Physics}\ }\textbf {\bibinfo {volume} {7}},\ \bibinfo {pages} {983} (\bibinfo {year} {2011})}\BibitemShut {NoStop}%
\bibitem [{\citenamefont {Ingber}(2003{\natexlab{a}})}]{ingber2003tensegrity1}%
  \BibitemOpen
  \bibfield  {author} {\bibinfo {author} {\bibfnamefont {D.~E.}\ \bibnamefont {Ingber}},\ }\bibfield  {title} {\bibinfo {title} {Tensegrity i. cell structure and hierarchical systems biology},\ }\href@noop {} {\bibfield  {journal} {\bibinfo  {journal} {Journal of cell science}\ }\textbf {\bibinfo {volume} {116}},\ \bibinfo {pages} {1157} (\bibinfo {year} {2003}{\natexlab{a}})}\BibitemShut {NoStop}%
\bibitem [{\citenamefont {Ingber}(2003{\natexlab{b}})}]{ingber2003tensegrity2}%
  \BibitemOpen
  \bibfield  {author} {\bibinfo {author} {\bibfnamefont {D.~E.}\ \bibnamefont {Ingber}},\ }\bibfield  {title} {\bibinfo {title} {Tensegrity ii. how structural networks influence cellular information processing networks},\ }\href@noop {} {\bibfield  {journal} {\bibinfo  {journal} {Journal of cell science}\ }\textbf {\bibinfo {volume} {116}},\ \bibinfo {pages} {1397} (\bibinfo {year} {2003}{\natexlab{b}})}\BibitemShut {NoStop}%
\bibitem [{\citenamefont {Chen}\ and\ \citenamefont {Ingber}(1999)}]{chen1999tensegrity}%
  \BibitemOpen
  \bibfield  {author} {\bibinfo {author} {\bibfnamefont {C.~S.}\ \bibnamefont {Chen}}\ and\ \bibinfo {author} {\bibfnamefont {D.~E.}\ \bibnamefont {Ingber}},\ }\bibfield  {title} {\bibinfo {title} {Tensegrity and mechanoregulation: from skeleton to cytoskeleton},\ }\href@noop {} {\bibfield  {journal} {\bibinfo  {journal} {Osteoarthritis and cartilage}\ }\textbf {\bibinfo {volume} {7}},\ \bibinfo {pages} {81} (\bibinfo {year} {1999})}\BibitemShut {NoStop}%
\bibitem [{\citenamefont {Bertoldi}\ \emph {et~al.}(2017)\citenamefont {Bertoldi}, \citenamefont {Vitelli}, \citenamefont {Christensen},\ and\ \citenamefont {Van~Hecke}}]{bertoldi2017flexible}%
  \BibitemOpen
  \bibfield  {author} {\bibinfo {author} {\bibfnamefont {K.}~\bibnamefont {Bertoldi}}, \bibinfo {author} {\bibfnamefont {V.}~\bibnamefont {Vitelli}}, \bibinfo {author} {\bibfnamefont {J.}~\bibnamefont {Christensen}},\ and\ \bibinfo {author} {\bibfnamefont {M.}~\bibnamefont {Van~Hecke}},\ }\bibfield  {title} {\bibinfo {title} {Flexible mechanical metamaterials},\ }\href@noop {} {\bibfield  {journal} {\bibinfo  {journal} {Nature Reviews Materials}\ }\textbf {\bibinfo {volume} {2}},\ \bibinfo {pages} {1} (\bibinfo {year} {2017})}\BibitemShut {NoStop}%
\bibitem [{\citenamefont {Rocklin}(2020)}]{rocklin2020flexible}%
  \BibitemOpen
  \bibfield  {author} {\bibinfo {author} {\bibfnamefont {D.~Z.}\ \bibnamefont {Rocklin}},\ }\bibinfo {title} {Flexible mechanical structures and their topologically protected deformations},\ in\ \href {https://doi.org/10.1007/978-3-642-27737-5_733-1} {\emph {\bibinfo {booktitle} {Encyclopedia of Complexity and Systems Science}}},\ \bibinfo {editor} {edited by\ \bibinfo {editor} {\bibfnamefont {R.~A.}\ \bibnamefont {Meyers}}}\ (\bibinfo  {publisher} {Springer Berlin Heidelberg},\ \bibinfo {address} {Berlin, Heidelberg},\ \bibinfo {year} {2020})\ pp.\ \bibinfo {pages} {1--16}\BibitemShut {NoStop}%
\bibitem [{\citenamefont {Liu}\ \emph {et~al.}(2022)\citenamefont {Liu}, \citenamefont {Bi}, \citenamefont {Yue}, \citenamefont {Wu}, \citenamefont {Yang},\ and\ \citenamefont {Li}}]{liu2022review}%
  \BibitemOpen
  \bibfield  {author} {\bibinfo {author} {\bibfnamefont {Y.}~\bibnamefont {Liu}}, \bibinfo {author} {\bibfnamefont {Q.}~\bibnamefont {Bi}}, \bibinfo {author} {\bibfnamefont {X.}~\bibnamefont {Yue}}, \bibinfo {author} {\bibfnamefont {J.}~\bibnamefont {Wu}}, \bibinfo {author} {\bibfnamefont {B.}~\bibnamefont {Yang}},\ and\ \bibinfo {author} {\bibfnamefont {Y.}~\bibnamefont {Li}},\ }\bibfield  {title} {\bibinfo {title} {A review on tensegrity structures-based robots},\ }\href@noop {} {\bibfield  {journal} {\bibinfo  {journal} {Mechanism and Machine Theory}\ }\textbf {\bibinfo {volume} {168}},\ \bibinfo {pages} {104571} (\bibinfo {year} {2022})}\BibitemShut {NoStop}%
\end{thebibliography}%
\clearpage
\section{Supplementary Material}

\preprint{APS/123-QED}

\subsection{A. System Description}
In this section, we explain the construction of the depleted triangular lattice and the characterization of its properties such as energy and rigidity probability using simulations. 
A periodic triangular lattice is constructed with basis vectors
\begin{align}
    b_1 &= (1, 0)\\
    b_2 &= (-0.5, \sqrt{3}/2)
\end{align}
The positions of the nodes are 
\begin{align}
    \{\mathbf{r_{\alpha\beta}}\}~=\{\alpha b_{1} + \beta b_{2}~|~\alpha, \beta ~\in [0, n]\}
\end{align}
where $n$ is the size of the system. 
In the system, there are $n^2$ nodes, each node surrounded by six neighbors. 
We apply periodic boundary conditions by restricting that the boundary nodes at $\mathbf{r}_{\alpha n}$ ($\mathbf{r}_{n \beta}$) connect to $\mathbf{r}_{\alpha 0}$ ($\mathbf{r}_{0 \beta}$), respectively. 
We must include the edge case that the node at $\mathbf{r}_{nn}$ connects only to the nodes at $\mathbf{r}_{0n}$, $\mathbf{r}_{n0}$, and $\mathbf{r}_{00}$. Disorder is introduced to the system by adding a random perturbation with magnitude between $-\delta$ and $\delta$ to $x$ and $y$ component of each node.
Suppose that we have the $i^{th}$ and $j^{th}$ node located at sites $\textbf{r}_i$ and $\textbf{r}_j$, we can find the unit vector pointing from the first node to the second as
\begin{align}
    \hat{\textbf{r}}_{ij} = \textbf{r}_{j} - \textbf{r}_{i}
\end{align}
A initial metric strain is applied to the system which deforms each bond in system giving the metric extension as
\begin{align}
    e_o &= \epsilon\cdot \hat{\textbf{r}}_{ij}
\end{align}
where $\epsilon$ is the metric deformation matrix defined as
\begin{align}
    \epsilon(\phi,~\theta) &= A(\theta)\epsilon_o(\phi)A(-\theta)\\
    \epsilon_o(\phi) &= \left(\begin{matrix}
                        \cos{\phi} + sin{\phi} & 0\\ 
                        0  & \cos{\phi} - \sin{\phi} \\ 
                        \end{matrix}\right)\\
  A(\theta) &= \left(\begin{matrix}
                        \cos{\theta} & -sin{\theta}\\ 
                        sin{\theta}  & \cos{\theta}\\ 
                        \end{matrix}\right)
\end{align}
We use the matrix $A(\theta)$ to average over isotropic strains over multiple runs.
The parameter $\phi$ controls the type of metric strain applied where $\phi = 0$ is dilation, $\phi = \pi/2$ is shear and $\phi = \pi$ is compression. 
The value of $\theta$ is determined by total number of runs, $\eta$
\begin{align}
    \theta = \frac{2\pi (k - 1)}{\eta}
\end{align}
where $k \in [1, \eta]$.

When the system relaxes, the two nodes connecting the bond (located at $\textbf{r}_i$ and $\textbf{r}_j$), undergo small displacements $\textbf{u}_i$ and $\textbf{u}_j$, respectively. The bond connecting the nodes undergoes an extension
\begin{align}
    e_{r} = \hat{\textbf{r}}_{ij}\cdot(\textbf{u}_j - \textbf{u}_i)
\end{align}
We can write this in form of the rigidity matrix $\mathbf{R}$ which relates the extensions of all the bonds in the system to the displacements of all the nodes as
\begin{align}
    \mathbf{e_{r}} = \textbf{R}\cdot\textbf{u}
\end{align}
where $\mathbf{e_{r}}$ is vector of bond extension due to relaxation, and $\mathbf{u}$ is the vector of node displacements. In the depleted triangular lattice system of size $n$, the rigidity matrix has dimensions $n_b\times 2n^2$ where $n_b \equiv 3n^2$ is the number of places where bonds can be present. Hence, we can write the total extension vector of all the bonds in the system due to metric strain and system relaxation as
\begin{align}
    \mathbf{e} &= \mathbf{e_{o}} + \mathbf{e_{r}}\\
    \mathbf{e} &= \mathbf{e_{o}} + \mathbf{R}\cdot\mathbf{u}
\end{align}
where $\mathbf{e_{o}}$ is vector of bond extensions due to metric strain. We can then write the energy of the system as considering the types of element present and the metric strain applied as 
\begin{align}\label{energy_matrix_form_sup}
    E = \frac{1}{2}\textbf{e}^{T}\cdot\textbf{K}(\textbf{e})\cdot\textbf{e}
\end{align}
where $\mathbf{K}(\mathbf{e})$ is the element matrix encoding the random positions of rods, cables, and struts. It is a $n_b \times n_b$ diagonal matrix with values 
\begin{align}
    K_{ii}(e_i) = \begin{cases}
        0             & \text{: no element}\\
        1             & \text{: rods}\\
        \Theta(e_i)   & \text{: cables}\\
        \Theta(-e_i)  & \text{: struts}
    \end{cases} 
\end{align}
for the $i^{th}$ element position in the system. The linear and non-linear elements are randomly placed in the system based on $\probrods$, $\probcables$, and $\probstruts$, which represent the fraction of bond locations in the system filled with rods, cables and struts, respectively. These parameters are used to construct $\mathbf{K}(\mathbf{e})$. We used \textit{Mathematica}'s FindMinimum function to minimize Eq. \ref{energy_matrix_form_sup}. We also define the fraction of bond locations that are non-linear elements as $\probNL \equiv \probcables+\probstruts$. For each value of $\probrods$, $\probcables$, $\probstruts$, the probability of rigidity $\probrigidity_\phi$ is measured by performing energy minimization over $\eta$ number of runs and the system is considered rigid if the energy is greater than 0. Numerical, $\probrigidity_{\phi}$ is obtained by
\begin{align}
    P_\phi = \frac{\textrm{number of times rigid}}{\eta}
\end{align}
In simulations, we are also able to vary the type of strain applied by changing $\phi$, and the ratio of non-linear elements present by varying $\nonlinearratio \equiv \probcables$/$\probNL$. 
The energy of the system can be also determined for each trial as a function of $\probrods$, $\probNL$, and $\nonlinearratio$ according to Eq. \ref{energy_matrix_form_sup}. 
\begin{table}[h]
    \centering
    \begin{tabular}{| c | c | c | c |}
        \hline
        $\probNL$ & $p_{crit}$ & $\alpha$ & $\beta$ \\
        \hline
        $0.0$ & $0.66$ & $1.100 \pm 0.015$ & $6.588 \pm 0.032$ \\
        \hline
        $0.1$ & $0.60$ & $1.478 \pm 0.070$ & $6.857 \pm 0.160$ \\
        \hline
        $0.2$ & $0.56$ & $1.582 \pm 0.085$ & $6.767 \pm 0.208$ \\
        \hline
        $0.3$ & $0.51$ & $1.806 \pm 0.130$ & $6.710 \pm 0.344 $\\
        \hline
    \end{tabular}
    \caption{The linear fit parameters of \ref{eq:energy_fit} to the energy $E$ of a system of rods and cables (without struts) undergoing compression, corresponding to the results shown in Fig. 2C of the main text.}
    \label{tab:fit_params_compression}
\end{table}
\begin{table}[h]
    \centering
    \begin{tabular}{| c | c | c | c |}
        \hline
        $\probNL$ & $p_{crit}$ & $\alpha$ & $\beta$ \\
        \hline
        $0.0$ & $0.65$ & $1.257 \pm 0.029$ & $6.773 \pm 0.061$\\
        \hline
        $0.1$ & $0.59$ & $1.033 \pm 0.032$ & $6.620 \pm 0.075$\\
        \hline
        $0.2$ & $0.53$ & $0.867 \pm 0.019$ & $6.528 \pm 0.046$\\
        \hline
        $0.3$ & $0.46$ & $0.673 \pm 0.025$ & $6.363 \pm 0.063$\\
        \hline
    \end{tabular}
    \caption{The linear fit parameters of \ref{eq:energy_fit} to the energy $E$ of a system of rods and cables (without struts) undergoing dilation, corresponding to the results shown in Fig. 2D of the main text.}
    \label{tab:fit_params_dilation}
\end{table}
In Fig. 2C and Fig. 2D of the main text, we perform linear fits to the energy simulation data for $\nonlinearratio = 0.0$ and $\probNL \in \{0.0, 0.1, 0.2, 0.3\}$ using the functional form
\begin{equation}\label{eq:energy_fit}
\ln E = \alpha \ln(\probrods - p_{\mathrm{crit}}) + \beta
\end{equation}
where $p_{\mathrm{crit}}$ denotes the value of $\probrods$ at which $E$ first becomes positive, which allows to observe the power-law scaling near the transition. 
The linear fit parameters and $p_{crit}$ are shown in Table \ref{tab:fit_params_compression}, and Table \ref{tab:fit_params_dilation} for Fig. 2C and Fig. 2D, respectively.

\subsection{B. Single Node Rigidity}
In this section we derive the analytical solution for the probability that a single node is rigid. The set of combinations of rod, cable, and strut that add to three bonds attached to a node is 
\begin{equation}\label{eq:element_config}
    \mathbb{E} = \{ (n_r, n_c, n_s) : n_r + n_c + n_s = 3\ |\ n_r, n_c, n_s \in \mathbb{N} \}
\end{equation}
where $n_r$ is the number of rods, $n_c$ is the number of cables and $n_s$ is the number of struts.
The probability of having a particular configuration of $n_r$ rods, $n_c$ cables, and $n_s$ struts given probabilities $\probrods$, $\probcables$, and $\probstruts$ is
\begin{equation}
    q(n_r, n_c, n_s) = \frac{\probrods^{~n_r}\probcables^{~n_c}\probstruts^{~n_s}(1 - \probrods - \probcables - \probstruts)^{6 - n_r - n_c - n_s}}{n_r!n_c!n_s!(6-n_r-n_c-n_s)!}
\end{equation}
where $\probrods$ is the probability of rods, $\probcables$ is the probability of cables, and $\probstruts$ the probability of struts in the system.
In a three-coordinated node structure, there are 20 unique configurations in which we can place the three tensioned bonds. 
We split these configurations into two classes, $A \to (+++)/(---)$ and $B \to (+-+)/(-+-)$ where $+$ and $-$ represent the sign of the tension on the elements. 
There are 8 configurations of class $A$ and 12 configurations of class $B$. 
\begin{table}[H]
    \centering
    \begin{tabular}{| c | c | c |}
         \hline
     $n_r$ $n_c$ $n_s$& $\Bar{q}_{A}$ & $\Bar{q}_{B}$ \\
     \hline
    1 1 1& 2/3  & 0\\
     \hline
    0 1 2& 1/3  & 0\\
     \hline
    0 2 1& 1/3  & 0\\
     \hline
    1 0 2& 1/3  & 1\\
     \hline
    1 2 0& 1/3  & 1\\
     \hline
    2 0 1& 1    & 1\\
     \hline
    2 1 0& 1    & 1\\
     \hline
    3 0 0& 1    & 1\\
     \hline
    0 3 0& 0    & 1\\
     \hline
    0 0 3& 0    & 1\\
     \hline
    \end{tabular}
    \caption{The probability that Class A, $\Bar{q}_{A}$, and Class B, $\Bar{q}_{B}$, configurations are rigid for different combination of $n_r$, $n_c$, and $n_s$ in a three-coordinated node configuration.}
    \label{tab:class_probability}
\end{table}
\noindent
We define the probability that the class A and class B configurations are rigid given $n_r$, $n_c$, and $n_s$ as $\Bar{q}_{A}(n_r, n_c, n_s)$ and $\Bar{q}_{B}(n_r, n_c, n_s)$, respectively. 
These probabilities are obtained by computing the fraction of configurations within each class that are force-balanced.
We then get the probability that a single three-coordinated node is rigid given $\probrods$, $\probcables$, and $\probstruts$ as
\begin{equation}\label{eq:single_node_probability}
    P_{N}(\probrods, \probcables, \probstruts) = \frac{1}{M}\sum_{\mathbb{E}} q(n_r, n_c, n_s) \left[\frac{3}{5}\Bar{q}_{A} + \frac{2}{5}\Bar{q}_{B}\right]
\end{equation}
where $M = \sum_{\mathbb{E}} q(n_r, n_c, n_s)$ is the normalization factor and we sum over all configurations in set $\mathbb{E}$.
The values of $\Bar{q}_{A}(n_r, n_c, n_s)$ and $\Bar{q}_{B}(n_r, n_c, n_s)$ are shown in Table \ref{tab:class_probability} for the different combinations of $n_r$, $n_c$, $n_s$. To test the analytical calculation, we simulate a small sub-network extracted from a larger disordered two-dimensional lattice, consisting of six nodes surrounding a central node.
In each of $N_{\rm trials}$, three edges connected to the central node are selected at random from a predefined set of six possible edges.
The sub-matrix of the rigidity matrix $\mathbf{R}$ corresponding to the selected edges is constructed, and its null space is computed to obtain the self-stress mode.
This mode is perturbed according to probabilities $\probrods$, $\probcables$, and $\probstruts$ by assigning each edge as a rod, a cable, or a strut.
For each perturbed configuration, the net force on the central node is calculated from stress values and the unit direction vectors of the edges.
The simulation measures the fraction trials in which rigid force-balanced configurations occur, yielding an empirical estimate of $P_{N}$.
The uncertainty in this estimate is calculated using the Wilson score confidence interval, which accounts for finite sample size effects.

\end{document}